\documentclass[aps,prr,twocolumn,superscriptaddress,preprintnumbers,showkeys]{revtex4-2}

\usepackage{amsmath,amssymb,amsfonts}
\usepackage{bm}
\usepackage{graphicx,color}
\usepackage{verbatim}
\usepackage{float}
\usepackage{dcolumn}
\usepackage{natbib}
\usepackage{hyperref}

\begin{document}

\title{Bacterial protein interaction networks: \\ connectivity is ruled by gene conservation, essentiality and function}

\author{Maddalena Dilucca}
\affiliation{Physics Department, Sapienza University of Rome, 00185 Rome (Italy)}
\author{Giulio Cimini}
\affiliation{Physics Department and INFN, University of Rome Tor Vergata, 00133 Rome (Italy)}
\affiliation{Institute for Complex Systems (CNR) UoS Sapienza, 00185 Rome (Italy)}
\author{Andrea Giansanti}
\affiliation{Physics Department, Sapienza University of Rome, 00185 Rome (Italy)}
\affiliation{INFN Roma1 unit, Sapienza University of Rome, 00185 Rome (Italy)}

%\date{\today}

\begin{abstract}
Protein-protein interaction (PPI) networks are the backbone of all processes in living cells. In this work we relate conservation, essentiality and functional repertoire of a gene to the connectivity $k$ (i.e., the number of interaction links) of the corresponding protein in the PPI network. 
On a set of 42 bacterial genomes of different sizes, and with reasonably separated evolutionary trajectories, we investigate three issues: i) whether the distribution of connectivities changes between PPI subnetworks 
of essential and nonessential genes; ii) how gene conservation, measured both by the evolutionary retention index (ERI) and by evolutionary pressures, is related to the the connectivity of the corresponding protein; iii) how PPI connectivities are modulated by evolutionary and functional relationships, as represented by the Clusters of Orthologous Genes (COGs). 
We show that conservation, essentiality and functional specialisation of genes constrain the connectivity of the corresponding proteins in bacterial PPI networks. In particular, we isolate a core of highly connected proteins (with connectivities $k\ge40$), which is ubiquitous among the species considered here -- though mostly visible in the degree distributions of bacteria with small genomes (less than 1000 genes). The genes that belong to this highly connected core are conserved, essential and, in most cases, belong to the COG cluster J, related to ribosomal functions and to the processing of genetic information.
\end{abstract}
\keywords{Protein-protein interactions, Gene Essentiality, Evolutionary Retention Index, Clusters of Orthologous Genes}

\maketitle

\section{Introduction}
\label{intro}

To operate biological activities in living cells, proteins work in association with other proteins, often assembled in large complexes. 
Hence, knowing the interactions of a protein is important to understand its cellular functions. Moreover, a comprehensive description of the stable and transient protein-protein interactions (PPIs) 
within a cell would facilitate the functional annotation of all gene products, and provide insight into the higher-order organisation of the proteome \cite{Drewes2003,Golemis2005}. 
Several methodologies have been developed to detect PPIs, and have been adapted to chart interactions at the proteome-wide scale. 
These methods, combining different technologies, experiments and computational analyses, generate PPI networks of sufficient reliability, 
enabling the assignment of several proteins to functional categories \cite{VonMering2002,Tong2002}. Moreover, the statistical study of bacterial PPIs over several species (meta-interactomes) has brought important knowledge about protein functions and cellular processes \cite{Shatsky2016,Caufield2017}. 

Our aim here is to shed some light on the relationships among conservation, essentiality and functional annotation at the genetic level and connectivities of PPI networks, at the protein level. We extend here our previous observations made on the PPI of \emph{E.coli} which suggested a strong correlation between the connectivity of PPI networks on the one hand, 
and codon bias, gene conservation and essentiality on the other hand \cite{Dilucca2015,Dilucca2018}. It is worth, in the next two paragraphs, specifying what is usually meant by gene essentiality and gene conservation.

Individual genes in the genome contribute differently to the survival of an organism. According to their known functional profiles 
and based on experimental evidence, genes can be divided into two categories: essential, and nonessential ones \cite{Gerdes2003,Fang2005}. 
Essential genes are not dispensable for the survival of an organism in the environment it lives in \cite{Fang2005,Peng2014}.
Nonessential genes are instead those which are dispensable \cite{Lin2010}, being related to functions that can be silenced without compromising the survival of the organism. 
Naturally, each species has adapted to one or more evolving environments and, plausibly, genes that are essential for one species may be not essential for another one.

It has been argued many times that essential genes are more conserved than nonessential ones \cite{Hurst1999,Jordan2002,Luo2015,Ish-Am2015,Alvarez-Ponce2016}. 
The term "conservation'' has, however, at least two meanings. On the one hand, a gene is conserved if orthologous copies of it are found in the genomes of many species, as measured by the Evolutionary Retention Index (ERI) \cite{Gerdes2003,Bergmiller2012}. 
On the other hand, a gene is (evolutionarily) conserved when it is subject to a purifying, selective, evolutionary pressure, which disfavours mutations.
A common measure of evolutionary pressure is $K_a/K_s$, the ratio of the number of non synonymous substitutions per non synonymous site to the number of synonymous substitutions per synonymous site. In this second meaning a conserved gene is, in a nutshell, a slowly evolving gene, a gene that hardly incorporates mutations \cite{Hurst1999,Hurst2002}. To measure the evolutionary pressures exerted on the genes we use here $K_a/K_s$, and to measure evolutionary patterns of codon bias we use the Effective Number of Codons (ENC) plots.

The main finding of this work is the presence, in bacterial PPI networks, of a functional transition ruled by the connectivity (degree $k$) of proteins. The genes of proteins with high connectivities are under selective pressure, conserved,  and essential. Below the transition ($k<50$), the functional repertoire of low connectivity proteins is heterogeneous, whereas the genes of proteins with $k>50$ mainly belong to the Cluster of Orthologous Genes (COG) J (related to translation, ribosomal structure and biogenesis), with just a few interesting hubs belonging to COGs I (Lipid transport and metabolism), K (Transcription) and L (Replication, recombination and repair). Moreover, we show that in the degree distribution of each bacterial PPI network there is an ubiquitous trace of an almost-invariant structure of conserved hubs, essentially due to the ribosomal protein complexes, mostly visible in the networks of bacteria with small genomes.

\section*{Materials and Methods}

\subsection*{\bf Bacterial dataset and PPI networks}

We consider a set of 42 bacterial genomes (that we have previously investigated in \cite{Dilucca2018}), here collected in Table~\ref{tab.dataset}. 
Nucleotide sequences were downloaded from the FTP server of the National Center for Biotechnology Information \cite{Benson2012}.
These genomes were chosen in order to have a reasonably large coverage of data concerning conservation, essentiality and selective pressure.

PPIs are obtained from the STRING database (Known and Predicted Protein-Protein Interactions, \url{https://string-db.org/})\cite{Szklarczyk2017}. We have chosen STRING because of its quite large coverage of different bacterial species, useful to extend to multiple species the study we did in \cite{Dilucca2015}. In STRING, each interaction is assigned with a confidence level or probability $w$, evaluated by comparing predictions obtained by different techniques \cite{Chien1991,Phizicky1995,Puig2001} with a set of reference associations, namely the functional groups of KEGG (Kyoto Encyclopedia of Genes and Genomes) \cite{Kanehisa2000}. In this way, interactions with high $w$ are likely to be true positives, whereas, a low $w$ possibly corresponds to a false positive. As usually done in the literature, we consider only interactions with $w\ge0.9$, a threshold that provides a fair balance between coverage and interaction reliability (see for instance the case study on $E.coli$ reported in \cite{Dilucca2015}). 
We denote by $k$ the \emph{degree} (number of connections) associated to each protein in each PPI network after the thresholding procedure. 
Note also that after applying the cut-off we are left, for each network, with a number of isolated proteins (singletons, with no connections) that grows as $\sqrt{n}$ 
(where $n$ is the number of proteins in the genome). These isolated proteins are not considered in the network analysis and are regarded as stemming from statistical noise or just appear isolated because the PPI data is incomplete.

It is known that PPIs of some species in our dataset might be known much better than others (this is for instance the case of $E.Coli$). To investigate potential bias in the dataset, we checked that the densities of PPIs are high for small genomes and tend to be constant and not so different from that of $E.coli$ in bacteria with bigger genomes (see bottom panel of Figure \ref{fig5b}). Moreover, the big genomes in our dataset include highly investigated pathogens.

The distinction between small and big genomes is a key emergent point in this work. We divide the set of 42 bacterial genomes in three groups, according to the number $n$ of their genes: a) $n < 1000$, b) $1000 < n < 3000$ and c) $n > 3000$. In the Supplementary Information we have addressed the dependence of various network properties on the size of the genome.

\subsection*{\bf Gene Conservation: ERI and $\mathbf{K_a/K_s}$} 

The Evolutionary Retention Index (ERI)~\cite{Gerdes2003} is a way of measuring the degree of conservation of a gene. 
In the present study the ERI of a gene is the fraction of genomes, among those reported in Table~\ref{tab.dataset}, that have at least an orthologous (same COG label) of the given gene. 
Then, as reminded in the Introduction, a low ERI value is related to a gene which is rather specific, common to a small number of genomes; whereas high ERI is characteristic of highly shared, putatively universal  and essential genes.

We also consider another notion of gene conservation. Conserved genes are those which are subject to a purifying, conservative evolutionary pressure. 
To discriminate between genes subject to purifying selection and genes subject to positive selective Darwinian evolution, 
we use a classic but still widely used indicator, the ratio $K_a$/$K_s$ between the number of non synonymous substitutions per non synonymous site ($K_a$) and the number of synonymous substitutions per synonymous site ($K_s$) \cite{Hurst2002}. 
This parameter represents a straightforward and effective way of separating genes subject to purifying evolutionary selection 
($K_a/K_s<1$) from genes subject to positive selective Darwinian evolution 
($K_a/K_s>1$). There are different methods to evaluate this ratio, though the 
alternative approaches are quite consistent among themselves. For the sake of 
comparison, we have used here the $K_a/K_s$ estimates by Luo et al. \cite{Luo2015} 
that are based on the Nej and Gojobori method~\cite{Nei1986}. 
Note that each genome has a specific average level of $K_a$/$K_s$ \cite{Dilucca2015}.

\subsection*{\bf Gene Essentiality} 
We used the Database of Essential Genes (DEG, \url{www.essentialgene.org}) \cite{Luo2015}, 
which classifies a gene as either essential or nonessential, on the basis of a combination of experimental evidence (null mutations or transposons) and general functional considerations. 
DEG collects genomes from Bacteria, Archaea and Eukarya, with different degrees of coverage\cite{Zhang2009,Zhang2014}. 
Of the 42 bacterial genomes we consider, only 23 are covered---in toto or partially---by DEG, as indicated in Table~\ref{tab.dataset}.

\subsection*{\bf ENC plot} 
The ENC-plot is a well known tool to investigate the patterns of synonymous codon usage in which the $ENC$ (Effective Number of Codons) values are plotted against $GC_3$  (Guanine and Cytosine Content at the third codon position). The formula of $ENC$ values expected under the hypothesis of pure mutational bias (no selection) is given by:
\begin{equation}
ENC =2+ s + \frac{29}{s^2+ (1-s)^2}
\label{eq1}
\end{equation} 
where $s$ represents the value of $GC_3$  \cite{Wright1990}. When the corresponding points fall near the expected neutral curve, mutations that enforce the typical mutational bias of the species are the main factor affecting the observed codon diversity. Whereas when the corresponding points fall considerably below the expected curve, the observed codon usage bias of the species is mainly affected by natural selection. To quantitatively represent the balance between mutational bias and selective natural pressure we parametrise the ENC formula, to be used in non-linear fits to the experimental data: 

\begin{equation}
ENC = a+ b*s + \frac{c}{s^2+ d* (1-s)^2}.
\label{eq2}
\end{equation} 
ENC plots of genes corresponding to low, intermediate and high connectivity proteins are shown in Figure \ref{fig4}. The best fit parameters for the three groups of genes are collected in Table \ref{tab5}.

\subsection*{\bf Clusters of orthologous proteins}
We use the functional annotation given in the database of orthologous groups of proteins (COGs) from Koonin's group, available at \url{http://ncbi.nlm.nih.gov/COG/} \cite{Tatusov2001,Galperin2015}.
We consider 15 functional COG categories (see Table~\ref{tab.func}), excluding the generic categories R and S for which functional annotation is either too general or missing. 

\section*{Results and discussion}

\begin{figure}[t!]
\centering
\includegraphics[width=0.49\textwidth]{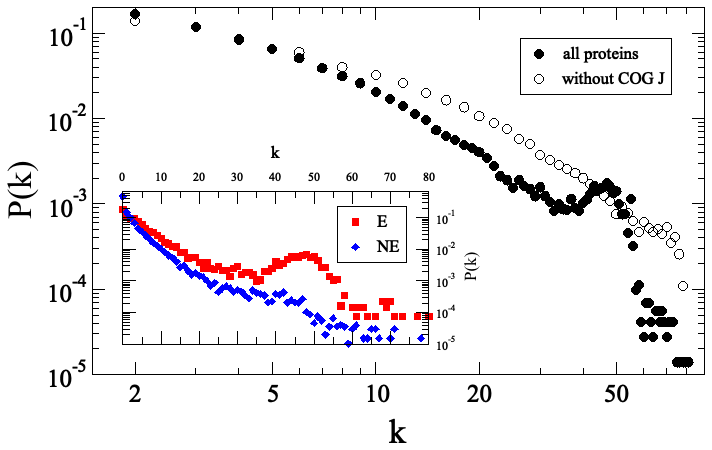}
\caption{Probability distribution $P(k)$ for the number of connections $k$ of each protein, averaged over the bacterial species considered in Table~\ref{tab.dataset} (full dots), compared with the degree distribution after removal of the proteins corresponding to genes in COG J, related to translational processes (empty dots). 
Inset: $P(k)$ for essential (E) and nonessential (NE) genes, averaged over DEG-annotated genomes. Note that the average degree is higher for essential genes than for nonessential ones, 
and the two probability distributions are quite distinct. The region of the curve for low $k$ can be well approximated by a power law \cite{Annibale2015}.} \label{fig1}
\end{figure}

\paragraph*{Degree distribution of PPI networks.}
We start by studying the degree distributions $P(k)$ observed in bacterial PPIs. We first recall that such a distribution was found to be scale-free in \emph {E.coli} 
\cite{Butland2005,Jin2013,Seesandra2014,Dilucca2015}, meaning that the corresponding PPI network features a large number of poorly connected proteins, and a relatively small number of highly connected hubs. 
In order to assess the generality of this observation, we compute $P(k)$ for each genome in Table \ref{tab.dataset} (plots are reported in Figures \ref{figPk1} and \ref{figPk2}). 
Note that, despite the fact that PPI networks of different bacteria have different sizes and densities, their average connectivity and the support of their $P(k)$ are very similar. Thus, we can superpose all the considered bacterial degree distributions without the need to normalise the support of each $P(k)$. 
When doing so, we observe two distinct regimes (see Figure \ref {fig1}). For low values of $k<$ 40, the distribution is approximately scale-free: $P(k)\propto k^{-\gamma}$ (${\gamma}=2.48$). 
This scaling behaviour is consistent with previous studies on the genomes of yeast, worms and flies \cite{Hahn2005} and on co-conserved PPIs in some bacteria \cite{Karimpour2008}. The scale free natura of bacterial PPI's is still a matter of debate, and a rough discussion of the origin of this feature is out of the scope of this paper. In this work we generally confirm that, as said above, there is as expected, a large number of poorly connected proteins and a small number of hubs. 

Remarkably, 
for higher values of $k$ the distribution deviates from a power law, and a bump with a Gaussian-like shape emerges.  This feature, visible for $k \ge 40$ may be due to the contribution of proteins belonging to large complexes \cite{Wuchty2014}. From the whole set of observation presented in this paper, the bump in the $P(k)$ is due to the complex of ribosomal interactions. Indeed, if one recalculates the degree distribution of a dataset in which the ribosomal proteins are removed the bump is not present (see Figure \ref{fig1}, empty dots). Moreover, if we consider the separate contribution of essential and nonessential genes to the $P(k)$ (for DEG-annotated genomes), 
we see that the bump is present only in the degree distribution of essential genes. Note also that the degree distributions for essential and nonessential genes are well separated and the average degree is systematically higher for essential genes than for nonessential ones, consistently with previous findings \cite{Hahn2005}. Remarkably, we have shown in a previous paper \cite{Dilucca2018} that the number of essential genes in bacteria  is close to 500 and does not depend on the size of the genome.  To correctly interpret the emergence of the bump in the average $P(k)$ in Figure \ref{fig1} it is worth to point out the distinction between small and not so small genomes. In the small genomes almost all the genes are essential and among the essential genes those belonging to COG J (functions related to translation and  ribosomal structure and biogenesis) play a major and ubiquitous role.  In Figure \ref{fig6b} we have checked that the bump that emerges in Figure \ref{fig1} as a feature of essential and conserved genes, is quite visible in the $P(k)$ of small genomes, whereas seems to be confused in the case of bigger genomes. This might be interpreted as a dilution effect; in the networks of bigger genomes there are a lot of specific interactions besides the essential ones. Nevertheless, averaging $P(k)$ over small, intermediate and big genomes we still see the bump and interpret it as an emerging feature due to a core of highly connected proteins (connectivities $k\ge40$). 
From these considerations we can exclude that this bump, observed here for the first time, might  emerge just because that part of the PPI is much more investigated than other subnetworks: 
the bump is there because of the ribosome, and this happens for all bacteria.

%FIGURE 1

\begin{figure}[t!]
\centering
\includegraphics[width=0.49\textwidth]{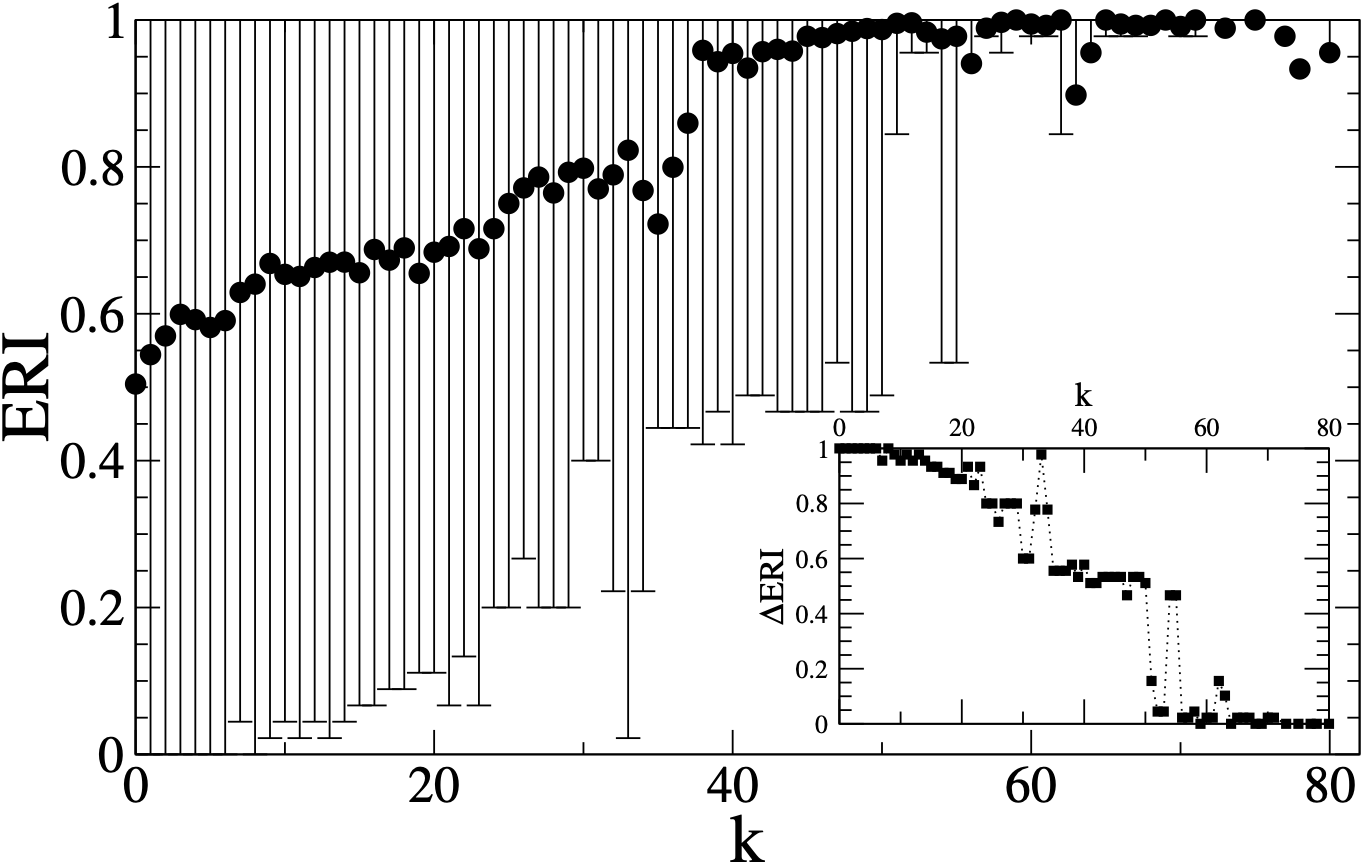}
\caption{Average ERI values of bacterial genes as a function of the degrees $k$ of the corresponding proteins, for all the considered genomes. 
Error bars are standard deviations of ERI values associated to a given $k$ value. 
Inset: amplitude of the error bar ($\Delta$ERI) as a function of $k$.}\label{fig2}
\end{figure}

\paragraph*{PPI connectivity and gene conservation}
We now investigate whether the connectivity $k$ of a protein in a PPI network drives a transition in the degree of conservation (as measured by ERI) of the corresponding genes. 
Figure \ref{fig2} displays the average value and the spread of ERI in genes relative to proteins with the same degree in the PPIs of different species. As a general feature we observe that, on the average, the genes of highly connected proteins are highly conserved among the bacterial species we consider, that constitute a reasonably wide sample of different evolutionary adaptations. The same Figure \ref{fig2} shows that if $k \le 50$ then the ERI highly fluctuates between different samples of proteins with the same $k$, in different species. For high connectivities (above $k=50$), the ERI is close to 1, with a drastic drop in the fluctuation (as shown in the inset). 
This observation points to the existence, in each bacterial PPI, of an almost-invariant structure of conserved hubs, sustained by highly conserved genes. We can conclude, as a rule of thumb, that a protein with connectivity degree of 40 or more is likely to be coded by a gene shared by at least 80$\%$ of the species in a generic pool of bacteria. At the moment, we have not a general explanation for this apparent threshold. Let us just  propose, as an heuristic observation, the existence of an almost-critical value of connectivity to be set between $40$ and $50$, that corresponds to the connectivity of the core of proteins specifically involved, as we have alluded to in the previous paragraph, to the ubiquitous ribosomal functions (see also Tables \ref{tab.hubs} and \ref{tab6}).

% FIGURE 2

\begin{figure}[t!]
\centering
\includegraphics[width=0.49\textwidth]{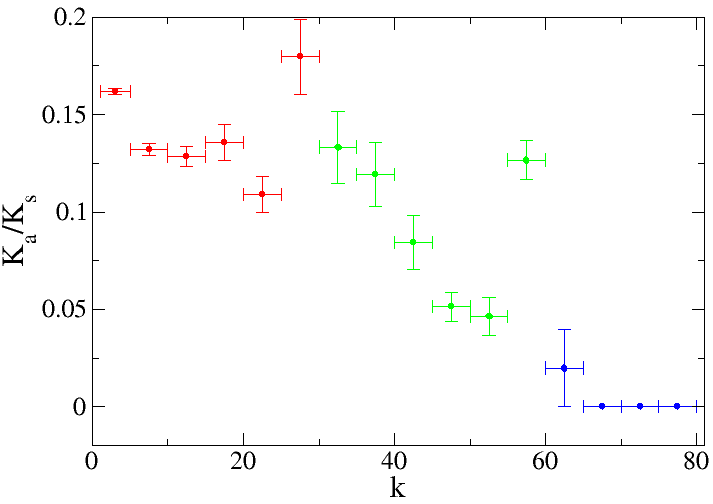}
\caption{($K_a/K_s$)  of groups of genes corresponding to proteins with different connectivity degrees $k$. As in the following Figures \ref{fig5} and \ref{fig4}, low connectivities are shown in red, intermediate in green and high connectivities in blue.} 
\label{fig3}
\end{figure}

\paragraph*{Evolutionary pressure and PPI connectivity}
We then look at the evolutionary pressure exerted on genes whose proteins have different connectivities.
Figure \ref{fig3} shows the ratio $K_a/K_s$ for groups of genes binned by the connectivity $k$ of the corresponding proteins, for all the 42 bacterial species in Table \ref{tab.dataset}. As is well known this ratio $K_a/K_s$ provides a straightforward indication of the balance between a positive driving {\it darwinian selection} (when the numerator prevails) and  a {\it purifying}, stabilising selection (acting against change in genes for which the denominator prevails).

We see that the more connected proteins correspond to genes which are subject to an increasing purifying evolutionary pressure. Indeed, the ratio ($K_a/K_s$) is less than 1 in all bins of connectivity and systematically decreases, as a function of $k$. A decreasing ratio generally indicates an increasing role of purifying, conservative, darwinian, evolutionary pressure on the corresponding set of genes. This is a reasonable results, pointing out that the groups of genes that support conserved structures of connectivity in the PPIs are more constrained, in evolution, than the genes of less interacting proteins.

% FIGURE 3

\begin{figure}[t!]
\centering
\includegraphics[width=0.49\textwidth]{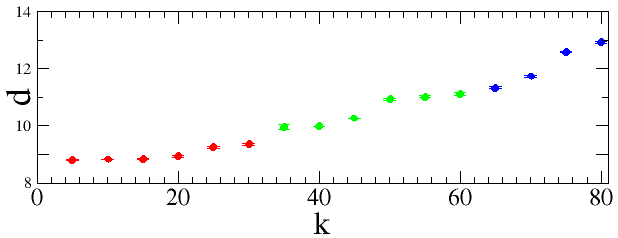}
\caption{ENC plot and connectivity. Each point in this graph represents a group of genes, characterised by the average connectivity $k$ of the corresponding proteins in the PPI network and by the average euclidean distance $d$, in the ENC plot, from Wright's theoretical curve. Different groups of genes are represented with different colors as a function of $k$. . The distance from the curve clearly increases with $k$. Wright's curve corresponds, in the ENC plot, to pure mutational bias (see equation (\ref{eq1})), then higher connectivities of the proteins imply bigger evolutionary selective pressure on the corresponding group of genes.}
\label{fig5}
\end{figure}

To add evidence to this observation we have also considered ENC plots for sets of genes binned by the connectivities of the corresponding proteins. Interestingly, the ENC data in Figure \ref{fig4} are fully consistent with those in Figure \ref{fig3}. In the ENC plots, the points associated to low connectivity proteins (red) are closer to the so called Wright's profile (represented there as black solid lines) than those associated to proteins with intermediate and high connectivities (green and blue lines).  Figure \ref{fig5} stresses this observation in a more quantitative way by showing that in the ENC plots the average distance from Wright's profile monotonously increases with $k$, Overall, the above results clearly indicate that codon bias and GC content of high connectivity genes are more under selective darwinian pressure than genes coding for low-connectivity proteins, in which the rate of accepted mutations is mainly ruled by neutral mutational bias. These observations point out that the almost-invariant structure of protein hubs we alluded to in the previous paragraph, is supported by and underlying set of genes which are under strong mutational control. Perhaps this is an expected result, but we clearly show it here as a general feature associated to ribosomal ubiquitous and conserved functions.

% FIGURE 5

\paragraph*{PPI and Essentiality.} 
To further investigate the relationship between gene essentiality and protein connectivities, we consider DEG-annotated genomes and 
classify interactions between proteins (links) making reference to the essentiality of the corresponding genes. We distinguish three sets of links: 
$ee$ (linking proteins from two essential genes), $\bar{e}\bar{e}$ (from two nonessential genes) and $e\bar{e}$ (from an essential gene and a nonessential one). 
We then compute the \emph{density} of these sets of links respectively as:

\begin{equation}\label{eq:sub1}
\rho_{ee}=\frac{|ee|}{\frac{1}{2}E(E-1)},
\end{equation}
\begin{equation}\label{eq:sub2}
\rho_{\bar{e}\bar{e}}=\frac{|\bar{e}\bar{e}|}{\frac{1}{2}NE(NE-1)},
\end{equation}
\begin{equation}\label{eq:sub2}
\rho_{e\bar{e}}=\frac{|e\bar{e}|}{\frac{1}{2}E\cdot NE},
\end{equation}
where $E$ and $NE$ denote the number of essential and nonessential genes, respectively (self-connections are excluded in our analysis). The denominator is the maximum possible value of the numerator, corresponding to the fully-connected graph.
Such densities are then compared with the overall density of the network---restricted to genes classified as either essential or nonessential:
\begin{equation}\label{eq:tot}
\langle\rho\rangle=\frac{|ee|+|\bar{e}\bar{e}|+|e\bar{e}|}{\frac{1}{2}(E+NE)(E+NE-1)}.
\end{equation}
We use the ratios $r_{ee}=\rho_{ee}/\langle\rho\rangle$, $r_{\bar{e}\bar{e}}=\rho_{\bar{e}\bar{e}}/\langle\rho\rangle$ and $r_{e\bar{e}}=\rho_{e\bar{e}}/\langle\rho\rangle$ 
to assess the level of connectivity of the subnetworks with respect to the overall connectivity. 
Table \ref{tab.rete} shows that subnetworks of essential genes are far denser than the overall networks, 
and that, in general, essential and nonessential genes tend to form network components that are weakly interconnected. 
This happens because many essential genes encode for ribosomal proteins, which in turn are localised in the ribosomal complex where they have a high probability to interact \cite{Bader2003} (see also Table 3 of \cite{Dilucca2018}, which shows approximately 25\% of essential genes fall into COG J).

Figures \ref{figM1} and \ref{figM2} display the superposed adjacency matrices of the $ee$ (red dots), $e\bar{e}$ (violet dots) and $\bar{e}\bar{e}$ (blue dots) subnetworks, thus showing the network features for each individual species. These graphs confirm the dominance of the interactions between the proteins of essential genes (red dots) in the small genomes. The adjacency matrices of bacteria with intermediate and big genomes are dominated by interactions involving proteins supported by non essential genes (blue dots).

\begin{figure}[t!]
\centering
\includegraphics[width=0.49\textwidth]{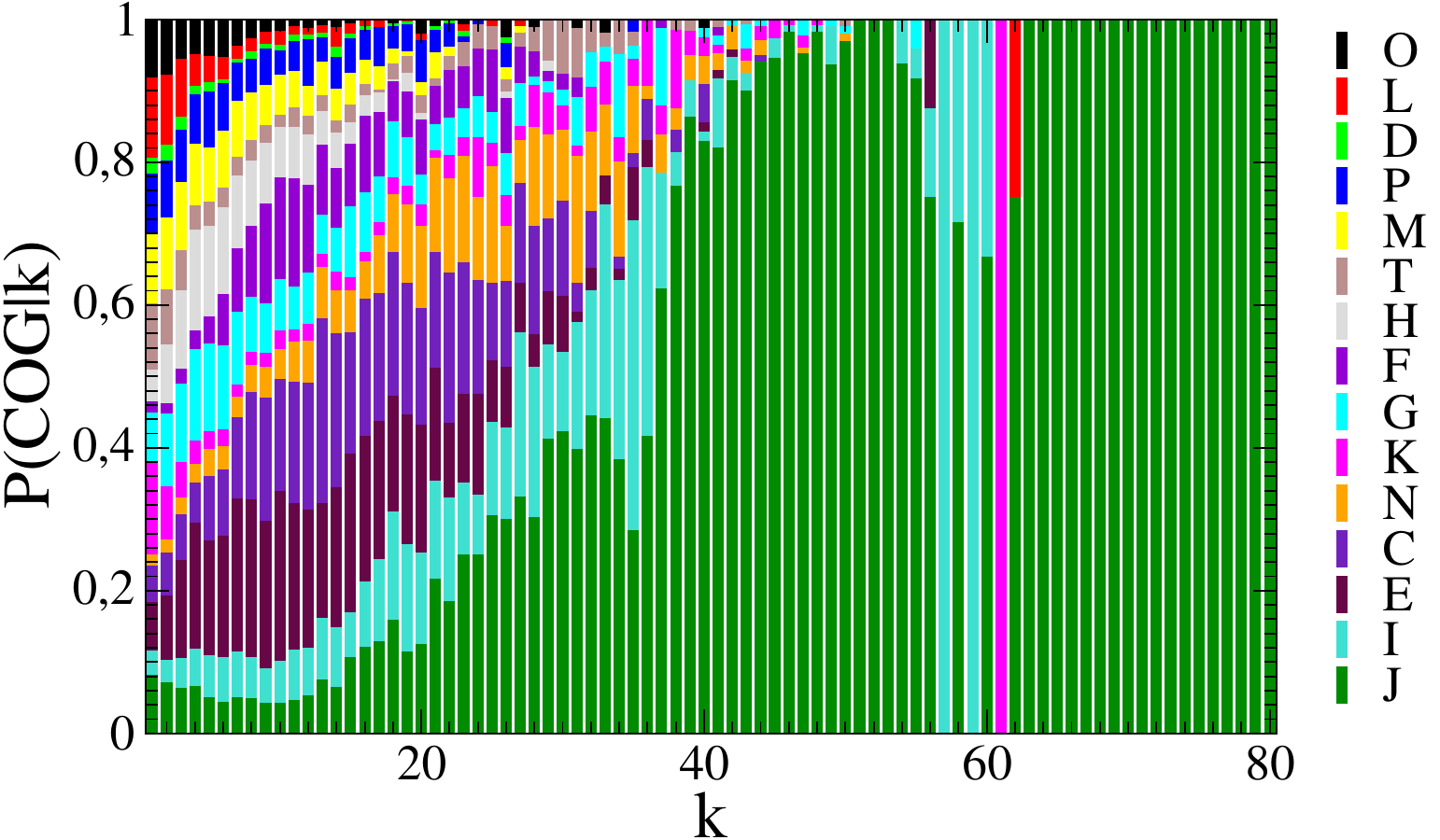}
\caption{Probability distribution $P(\mbox{COG}|k)$ of belonging to a given COG for proteins with degree $k$, over all considered genomes. 
Proteins with low connectivity have a very heterogeneous COG composition, whereas, those with high $k$ basically belong only to COG J.}\label{fig6}
\end{figure}

\paragraph*{PPI connectivity and functional specialisation.}
For each PPI network, we define the conditional probability that a protein with degree $k$ belongs to a given COG as: 
\begin{equation}\label{eq:pcog}
P(\mbox{COG}|k)= P(k|\mbox{COG})P(\mbox{COG})/P(k),
\end{equation}
where $P(k)$ is the degree distribution in the PPI network, $P(\mbox{COG})$ is the frequency of that COG in the proteome, and $P(k|\mbox{COG})$ is the degree distribution restricted to that COGs. 
Figure \ref {fig6} shows the COG spectrum as a function of $k$ over all the bacteria here considered. Interestingly, we again note a marked transition. 
Below $k \simeq 40$ the COG spectrum is quite heterogeneous: genes corresponding to proteins with low connectivity are spread over several COGs which correspond to different functions (see Table \ref{tab.func}). Instead, proteins with more than 40 interactions are likely to be coded by genes belonging to COG J. There are yet a handful of outliers, hubs with connectivities between 57 and 62, that belong to COG I (related to lipid transport and metabolism) 
and K and L (which, together with J, define the functional class of information storage and processing). The list of these outliers is reported in Table \ref{tab.hubs}. Interestingly, they correspond to RNA polymerases and to enzymes involved in the acetate metabolism. But, which are the genes of COG J that drive the transition? 

\begin{figure*}[t!]
\centering
\includegraphics[width=0.9\textwidth]{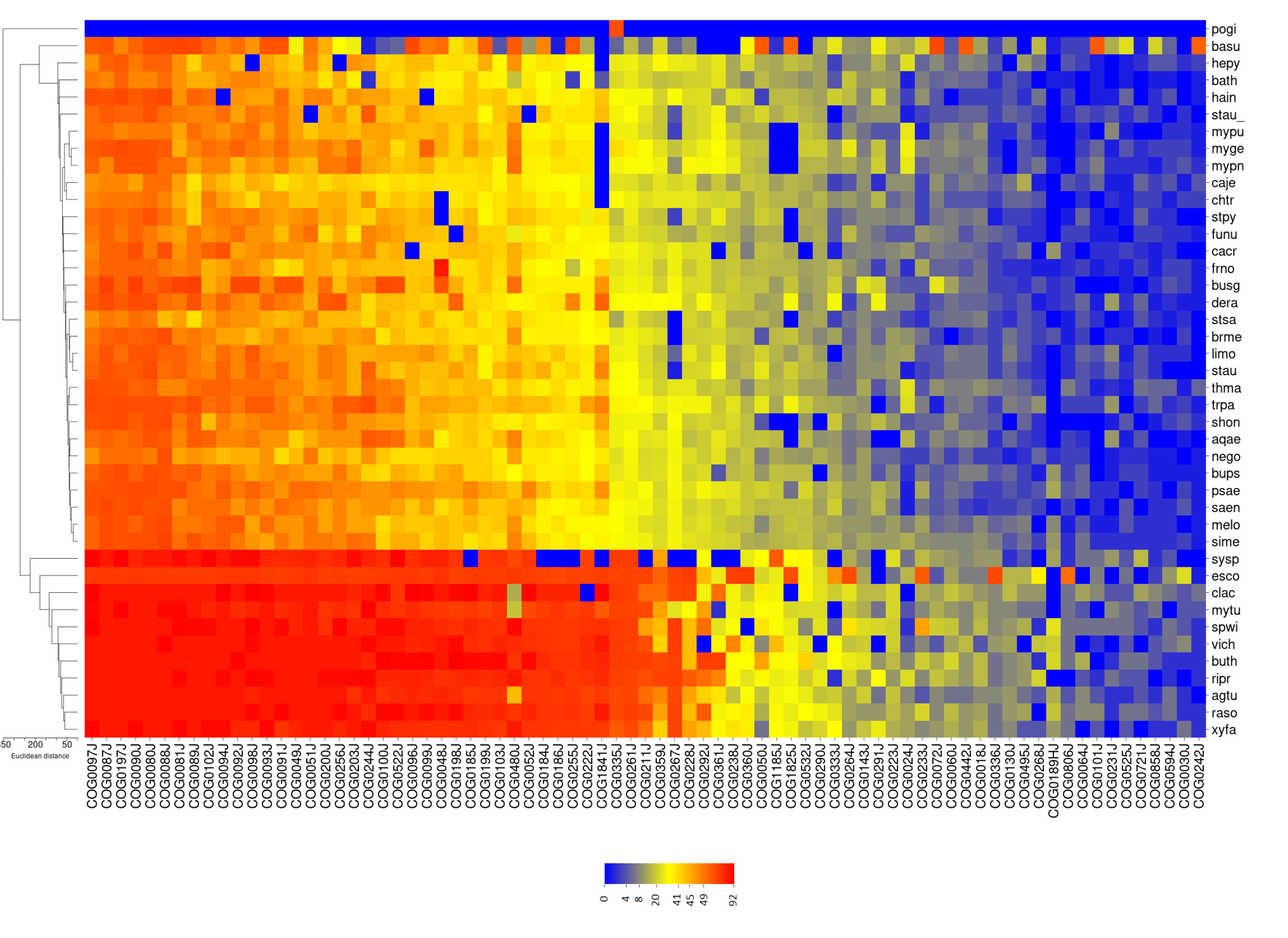}
\caption{Heat map of the connectivity degree of the protein as distributed over the COG J genes with ERI=1, in each species. Genes are sorted by decreasing average degree. We note that those genes which correspond to degrees bigger than 40 are conserved for all species. Details of these genes are provided in Table \ref{tab6}.}\label{fig7}
\end{figure*}

In Figure \ref{fig7} we are able to show which genes are the main characters in the transition. We investigate the connectivities of the highly conserved genes (ERI=1, shared by all the species in Table \ref{tab.dataset})  belonging to COG J, and whose proteins have connectivities bigger than $40$. These highly shared genes corresponding to cores of highly connected ribosomal proteins are listed in Table \ref{tab6}. In the heat map of Figure \ref{fig7} we sort each gene in the COG J in order of descending degree, species by species, and we see that there is a core of genes (in red, lower left sector) that correspond to highly connected proteins, which are also highly shared (ERI = 1, see Table \ref{tab6}) among all the species we considered.
It is quite clear from this heat mapthat the 42 species in this study can be split into at least two groups (see the cladogram on the left). In the bottom group there is a shared set of genes (the red band at the bottom-left side of the heat map) corresponding to a common core of highly connected ribosomal proteins. This remarkable observation suggests that the species in this group (namely, Synechocystis sp. PCC 6803, Escherichia Coli K-12 MG1655, Clostridium acetobutylicum ATCC 824, Mycobacterium tuberculosis H37Rv, Sphingomonas wittichii RW1, Vibrio cholerae N16961, Burkholderia thailandensis E264, Rickettsia prowazekii str. Madrid E, Agrobacterium tumefaciens (fabrum), Ralstonia solanacearum GMI1000, Xylella fastidiosa 9a5c) should have a common structural and functional organisation of their ribosomes -- An Interesting point to be further investigated. In the rest of the species the connectivity of the proteins, corresponding to the highly shared COG J genes, with $k>40$ is more heterogeneous. We can conclude that the abrupt transition shown in Figure \ref{fig6} is driven by a subset of COG J genes which are highly conserved among a subset of species and are listed in Table \ref{tab6}. As one can see these genes correspond to a specific subset of ribosomal proteins in the small and large subunits that should be further investigated in their functional and structural role.

% FIGURE 6

% FIGURE 7 

\section*{Conclusions}

Connectivity analysis of biological networks, such as protein-protein interaction or metabolic networks, 
has demonstrated that structural features of network subgraphs are correlated with biological functions \cite{Srinivasan2006,Rao2014}. 
For instance, it was shown that highly connected patterns of proteins in a PPI are fundamental to cell viability \cite{Jeong2001}.
In this work we have shown the existence of a functional transition  in bacterial species, ruled by the connectivity of proteins in the PPI networks. The critical threshold in $k$ of the transition is located between $k=40$ and $k=50$. Proteins that have connectivities above the threshold are mostly encoded by genes that are conserved, under selective pressure (as measured both by ERI and $K_a/K_s$) and essential. Moreover the functional repertoire above the threshold focuses mainly on the COG J (Translation, ribosomal structure and biogenesis), with just a few interesting hubs belonging to COGs I (Lipid transport and metabolism), K (Transcription) and L (Replication, recombination and repair).       

Indeed, the PPI network of each bacterial species is characterised by a highly connected core of conserved ribosomal proteins, 
the components of multi-subunit complexes whose corresponding genes are mostly essential \cite{Butland2005,Karimpour2008} and code for supra-molecular complexes, 
that pile up in the bump we have observed for the degree distribution (Figure \ref{fig1}). 
Hence, what we are seeing here is essentially the ribosome, and related protein complexes such as RNA Polymerase. 
Indeed, the ribosome is the only molecular machine in bacteria in which a given protein could legitimately have 40 or more protein binding partners, with the help of rRNA mediating interactions \cite{Fox2010}. 

Admittedly, since there are bacterial species that are much more investigated than others, comparative statistical studies of bacterial PPIs might be particularly biased by the choice of the sample of genomes to be included in the study. Our dataset is no exception. In order to assess this possible bias in our study we have checked that in our dataset we have included small genomes (i.e. less than 1000 genes) whose PPIs have densities (a rough proxy for the coverage of the interactions in the network) that are higher than those of bigger genomes (Figure \ref{fig5b}). 
The group of small genomes comprises Buchnera, Chlamydia, Mycoplasmas, whereas bigger genomes refer mostly to illustrious pathogens that are surely among the most investigated bacterial species. The densities of the networks of these species are quite similar and comparable with that of \emph {E.coli}. As a general rule, and quite obviously, the networks relative to small genomes are better covered in the STRING database (after the application of a conservative cutoff $w = 900$) than those relative to bigger genomes. Interestingly, we have shown that, indeed, the PPI adjacency matrices of bacteria with small genomes are dominated by the interactions constituting the ribosomal complex. In the adjacency matrices of the PPIs of bacteria with bigger genomes, the cloud of interactions between the proteins of non essential genes tends to superpose to the ever present ribosomal core.  In conclusion, we believe to have convincingly shown that bacterial PPIs are characterised by the presence of a highly connected structure, associated to the ribosomal functions, and particularly visible in bacteria with small genomes.

The observations we have presented here could be useful for the prediction of gene essentiality, based on the knowledge of PPI networks, and for the prediction of interactions between proteins, based on genetic information\cite{Hwang2009,Wei2013}.
It is interesting to note that our results are consistent with a previous study based on inferred bacterial co-conserved networks based on phylogenetic profiles \cite{Karimpour2008}. 
This work suggests to further and systematically investigate how the structure of the PPI networks is correlated with multiple networks at the genetic level, at least in unicellular organisms. In particular we believe that a recent approach based on the introduction of multiple-layer networks could be of great potential interest (e.g. to search for a general scheme behind antimicrobial resistance \cite{Bardini2018,Terradot2011,Zoraghi2013,Sevimoglu2014,Boccaletti2014}).

\newpage\newpage

\newcommand{\avg}[1]{\langle #1\rangle}

\renewcommand{\thefigure}{S\arabic{figure}}
%\setcounter{figure}{0}  

% FIGURE 1
\begin{figure*}[t!]
\centering
\includegraphics[width=0.49\textwidth]{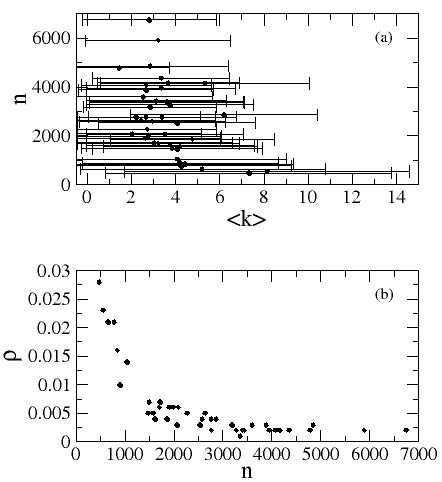}	
\caption{Relation between genome size $n$ and average degree $\avg{k}\pm\sigma_k$ (upper panel) and density $\rho$ (bottom panel) of the corresponding PPI network for the set of bacteria collected in Table \ref{tab.dataset}.The density of a network is the ratio between the actual number of links and the number of links in the fully connected case, namely $\frac{1}{2}n(n-1)$.}
\label{fig5b}
\end{figure*}

% FIGURE 2
\begin{figure*}[t!]
\centering
\includegraphics[width=0.49\textwidth]{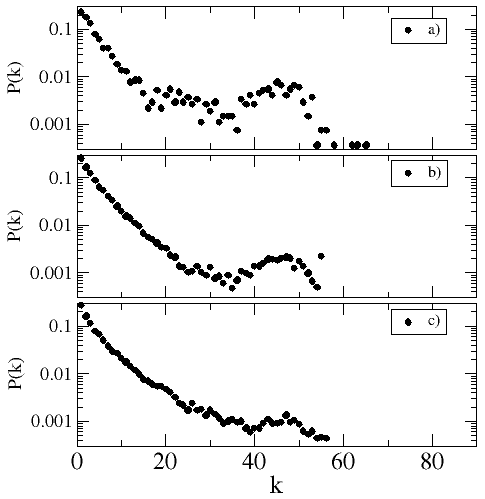}	
\caption{Distributions of PPI degree $P(k)$  averaged over the PPI networks of the three groups of bacteria in Table \ref{tab.dataset}, subdivided according to the number $n$ of their genes: a) $n <1000$, b) $1000< n <3000$ and c) $n >3000$. Clearly, the "bump" at $k>40$ in Figure \ref{fig1} is a feature that is characteristic of small genomes (groups a and b), mostly constituted by essential genes. In the genomes of group c the interactions that constitute the bump are diluted among the other interactions due to non essential genes. This dilution effect can be seen also in the subsequent figures \ref{figPk1}, \ref{figPk2}, \ref{figM1}, \ref{figM2}}
\label{fig6b}
\end{figure*}

% FIGURE 4
\begin{figure*}[h!]
\centering
\includegraphics[width=0.80\textwidth]{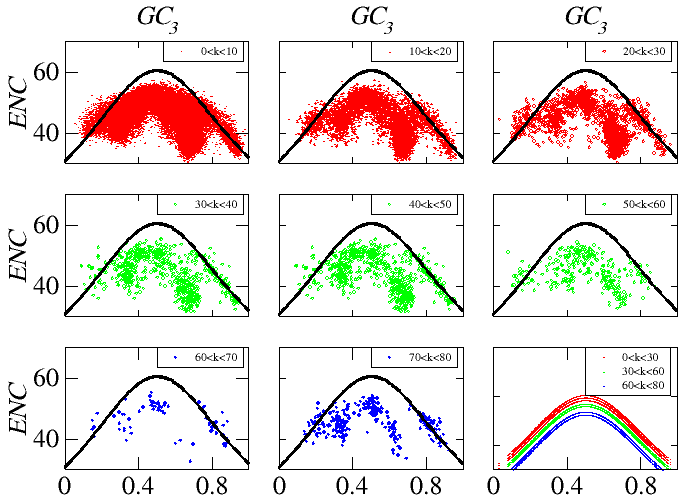}
\caption{ ENC plots for three groups of genes corresponding to proteins with different degree connectivities $k$. In each panel the solid black lines are plots of Wright's theoretical curve (equation (2)) which correlates effective number of codons with $GC_3$  in the case of pure mutational bias (no selective pressure).  Coherently with Figures \ref{fig3} and \ref{fig5}, the case of low connectivities are shown in red, intermediate in green and high connectivities in blue. In the bottom-right panel dashed non-linear fits of  Wright's theoretical shapes  to the experimental data. For the sake of completeness the best fit parameters are reported in the following Table \ref{tab5}.}\label{fig4}
\end{figure*}

\begin{figure*}[h!]
\centering
\includegraphics[width=\textwidth]{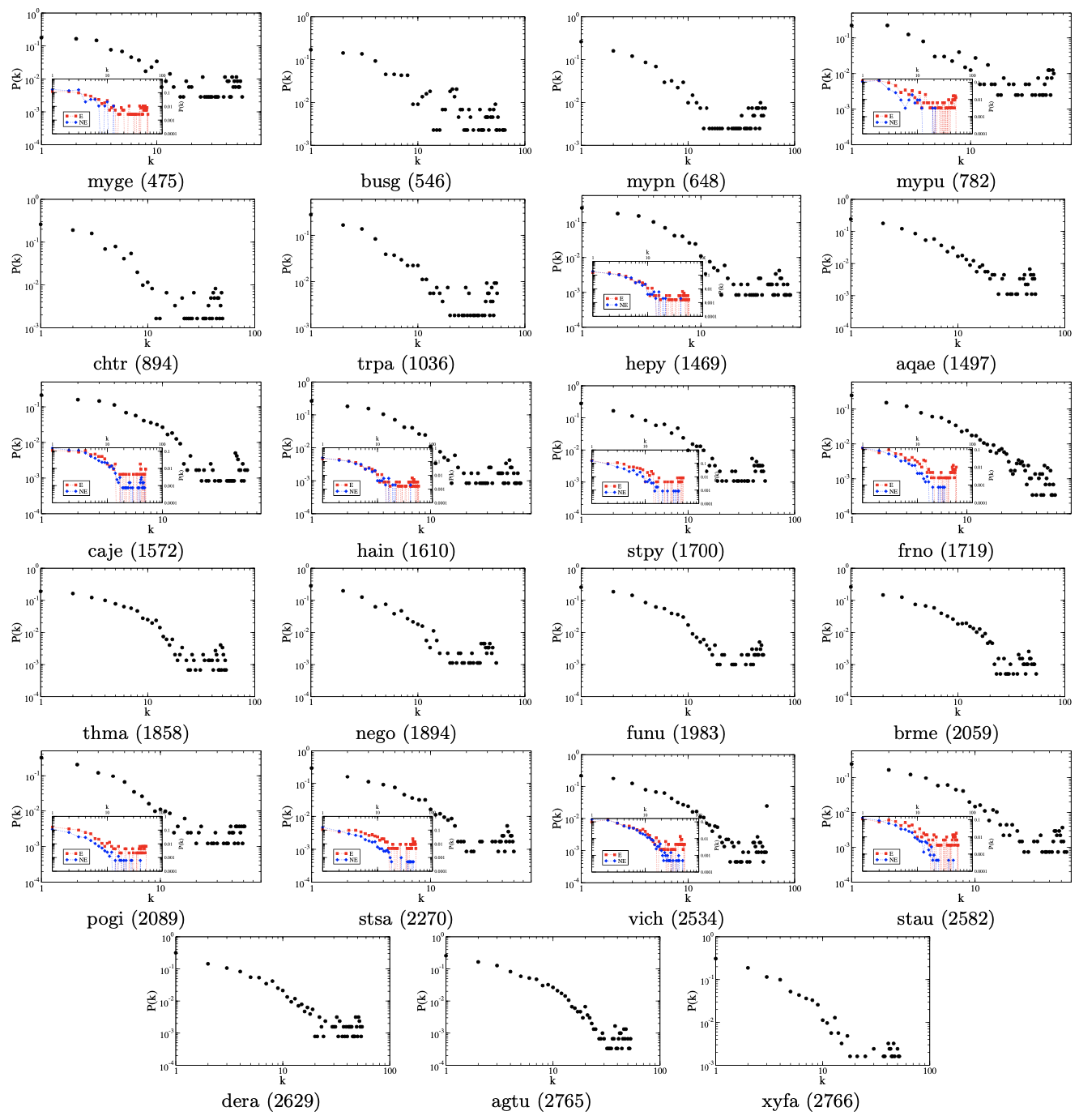}	
\caption{Degree distributions $P(k)$ (part 1, small genomes) of some of the bacteria collected in Table \ref{tab.dataset}, sorted by the increasing size n of their genomes. For DEG-annotated genomes, the inset shows the contribution of essential (red) and nonessential (blue) genes. Note, for $k>40$, in many cases the presence of a structure, particularly evident in the $P(k)$ of the essential genes, that likely contributes to the bump in Figure \ref{fig1}.}
\label{figPk1}
\end{figure*}

\begin{figure*}[h!]
\centering
\includegraphics[width=\textwidth]{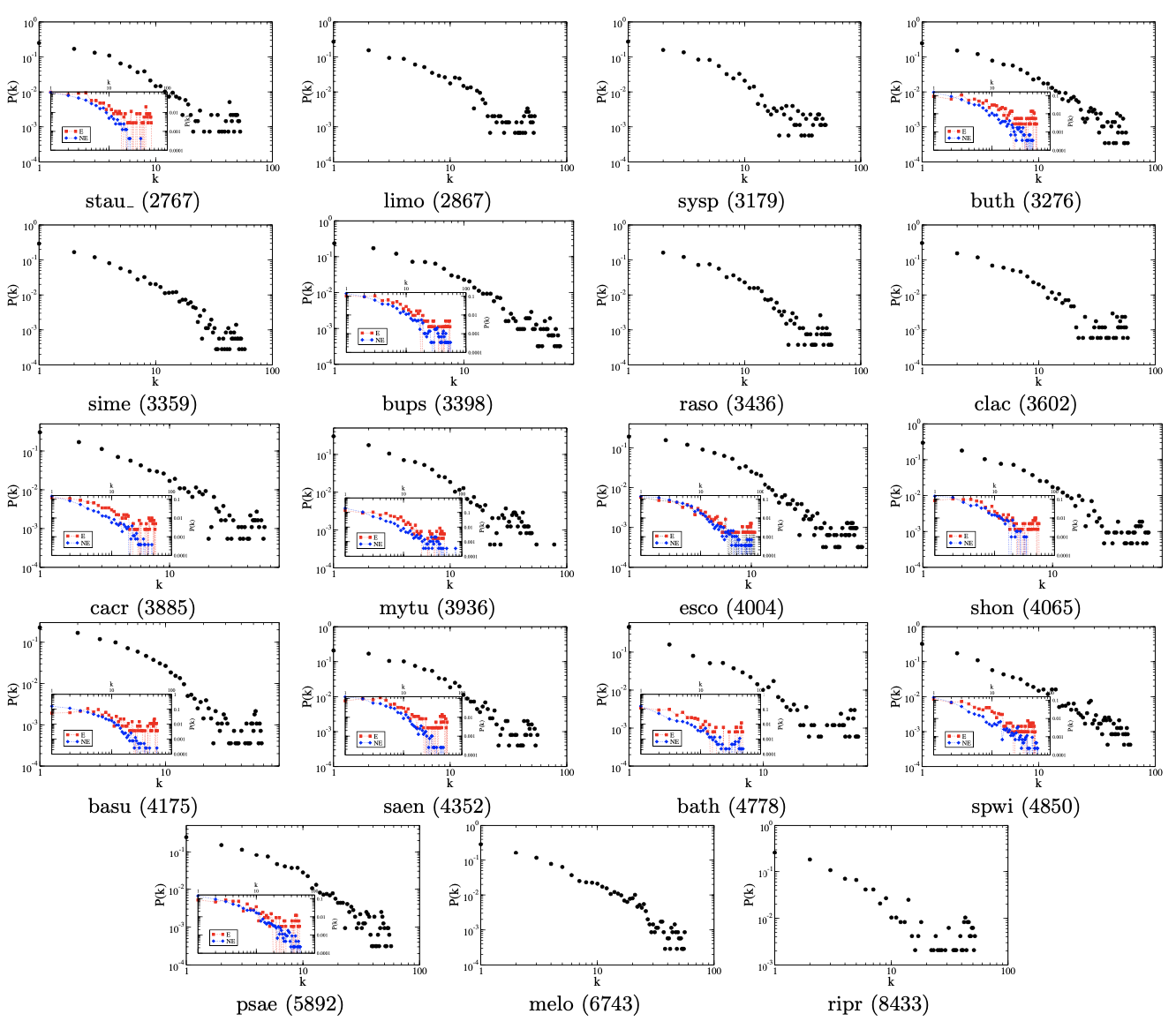}	
\caption{Degree distributions $P(k)$ (part 2, big genomes) of the PPI networks for some of the bacteria collected in Table \ref{tab.dataset}, ordered by the increasing size $n$ of their genomes. For DEG-annotated genomes, the inset shows the contribution of essential (red) and nonessential (blue) genes. Note that, in most cases, in the region $k>40$, the signature of the bump is blurred, hidden behind a general power law trend, likely due to the contribution of the interactions of non essential genes, as shown in the subsequent figures: \ref{figM1} and \ref{figM2}.}
\label{figPk2}
\end{figure*}

\begin{figure*}[h!]
\centering
\includegraphics[width=\textwidth]{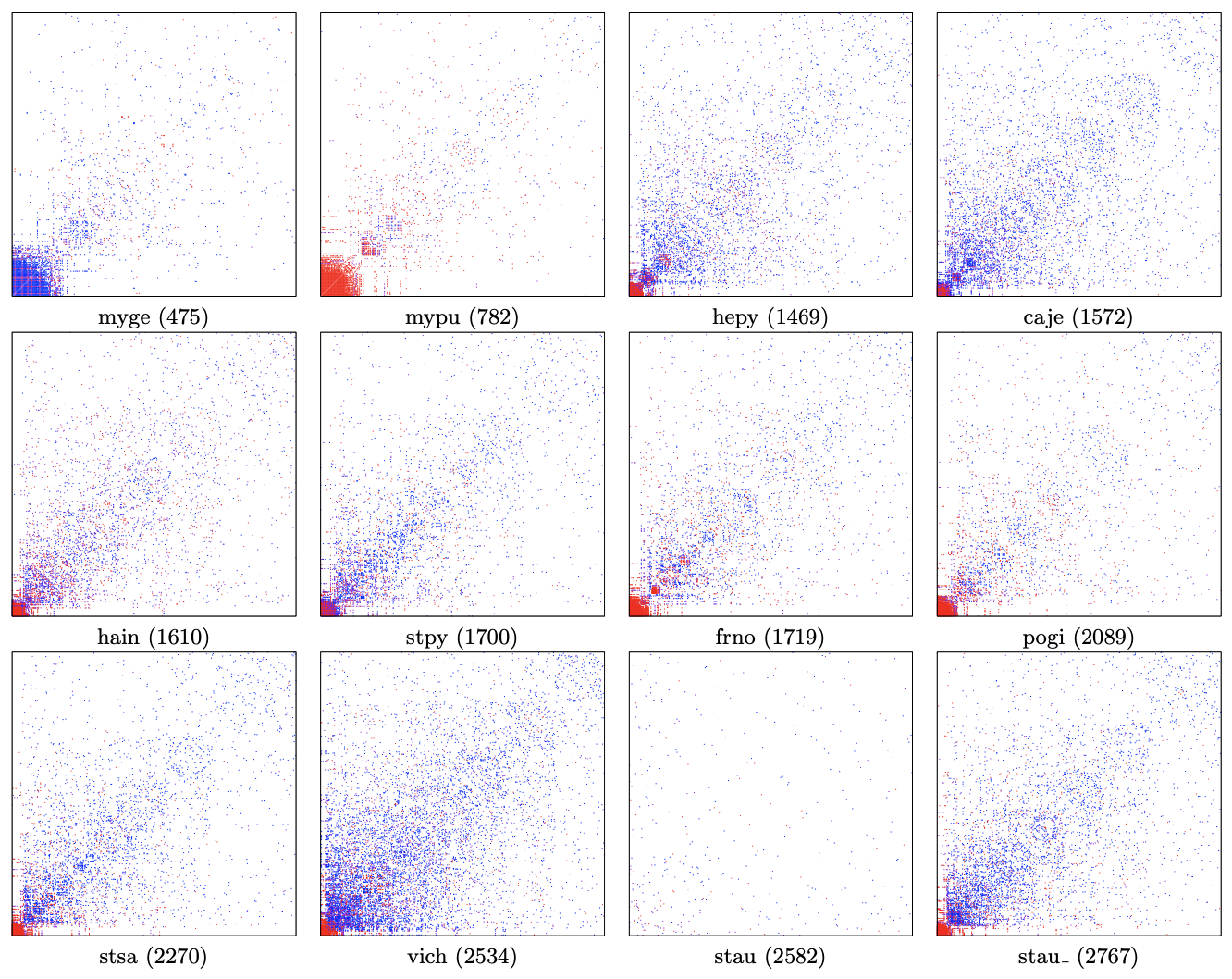}	
\caption{Adjacency matrices (part 1, small genomes) of PPI networks for the DEG-annotated bacterial species collected in Table \ref{tab.dataset}. From top left the matrices are sorted according to the increasing size of the bacterial genomes. In each matrix, genes are ordered according to the decreasing degree of the corresponding protein in the network, from left to right (horizontal axis) and  from bottom to top (vertical axis). Links between essential genes are plotted as a red dot, those between nonessential (and non-annotated) genes with blue dots, 
and those between essential and nonessential (plus non-annotated) genes with a violet dot. The red spot in the lower left sector of each matrix corresonds to the core of genes of the highly connected proteins of the almost-invariant structure of ribosomal hubs. Overall, in the case of small genomes it is evident how the red subnets of the essential genes dominate the matrices. In bigger genomes, as also shown in the next figure, the blue non-essential with non-essential gene  interactions and the violet cross-interactions tend to superpose to the core of the essential genes. It is worth noting that it could be reasonable to evaluate how much a PPI is relatively covered, in each species by the relative occurrence of blue and violet dots normalized to the maximal number of links in the network}
\label{figM1}
\end{figure*}

\begin{figure*}[h!]
\centering
\includegraphics[width=\textwidth]{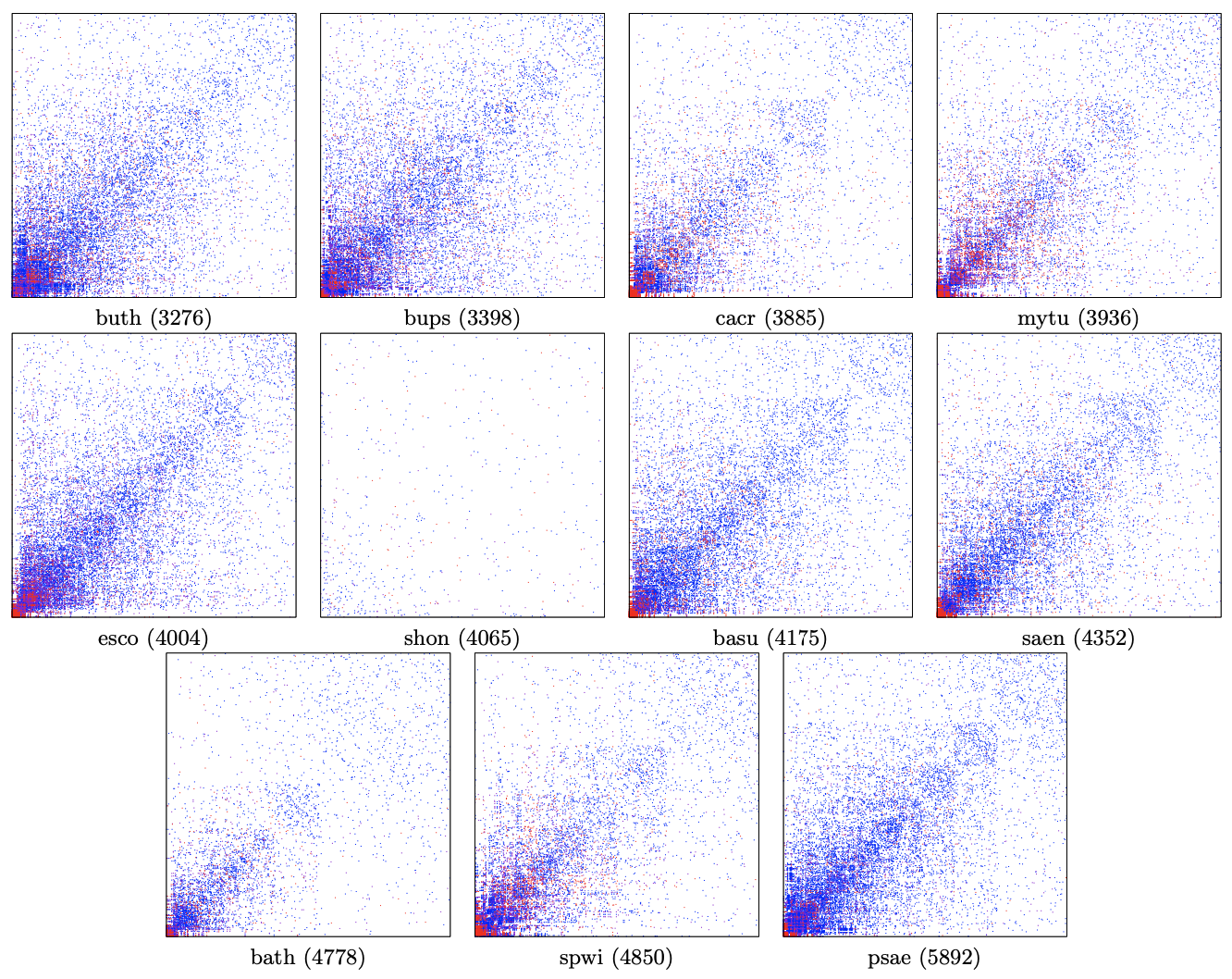}	
\caption{Adjacency matrices (part 2, big genomes) of PPI networks for the DEG-annotated bacterial species collected in Table \ref{tab.dataset}. From top left the matrices are sorted according to the increasing size of the bacterial genomes. In each matrix, genes are ordered according to the decreasing degree of the corresponding protein in the network, from left to right (horizontal axis) and  from bottom to top (vertical axis). Links between essential genes are plotted as a red dot, those between nonessential (and non-annotated) genes with blue dots, 
and those between essential and nonessential (plus non-annotated) genes with a violet dot. Also looking at these graphs one would be tempted to possibly associate the relative presence of blue interactions to the extent the interactome is known and annotated
}\label{figM2}
\end{figure*}

%TABLE 1(tab.dataset)
\begin{table*}[p]
\centering
\begin{tabular}{lccccc} 
\hline
{\bf Organisms}	&	{\bf Abbr.} &  {\bf Class} & {\bf RefSeq} &	{\bf STRING}  &$\mathbf{n}$  \\
\hline
\bf
Mycoplasma genitalium G37	&	myge	& 10&	NC\_000908	&	243273	&	475 \\
Buchnera aphidicola Sg uid57913	&	busg	&	2 & NC\_004061	&	198804	&	546 \\
Mycoplasma pneumoniae M129	&	mypn	& 10 &	NC\_000912.1	&	272634	&	648 \\
\bf
Mycoplasma pulmonis UAB CTIP	&	mypu	&	10 & NC\_002771	&	272635	&	782 \\
Chlamydia trachomatis D/UW-3/CX	&	chtr	& 14 &	NC\_000117.1	&	272561	&	894 \\
Treponema pallidum Nichols	&	trpa	&	11 &NC\_000919.1	&	243276	&	1036 \\
\bf
Helicobacter pylori 26695	&	hepy	&	4 & NC\_000915	&	85962	&	1469 \\
Aquifex aeolicus VF5	&	aqae	&	12 & NC \_000918	&	224324	&	1497 \\
\bf
Campylobacter jejuni 	&	caje	&	 4 & NC\_002163	&	192222	&	1572 \\
\bf
Haemophilus influenzae Rd KW20	&	hain	&	3 &NC\_000907.1	&	71421	&	1610 \\
\bf
Streptococcus pyogenes NZ131	&	stpy	&	6 &NC\_011375	&	471876	&	1700 \\
\bf
Francisella novicida U112	&	frno	&	3 & NC\_008601	&	401614	&	1719 \\
Thermotoga maritima MSB8	&	thma	&	16 &NC\_000853.1	&	243274	&	1858 \\
Neisseria gonorrhoeae FA 1090 uid57611	&	nego	& 2 & 	NC\_002946	&	242231	&	1894 \\
Fusobacterium nucleatum ATCC 25586	&	funu	&	15 &NC\_003454.1	&	190304	&	1983 \\
Brucella melitensis bv. 1 str. 16M	&	brme	&	1 & NC\_003317.1	&	224914	&	2059 \\
\bf
Porphyromonas gingivalis ATCC 33277	&	pogi	&	7 &NC\_010729	&	431947	&	2089 \\
\bf
Streptococcus sanguinis	&	stsa	&	6 &NC\_009009	&	388919	&	2270 \\
\bf
Vibrio cholerae N16961	&	vich	&	3 &NC\_002505	&	243277	&	2534 \\
\bf
Staphylococcus aureus N315	&	stau	& 6 &	NC\_002745.2	&	158879	&	2582 \\
Deinococcus radiodurans R1	&	dera	&	9 & NC\_001263.1	&	243230	&	2629 \\
Agrobacterium tumefaciens (fabrum)	&	agtu	& 1 &	NC\_003062 &	176299	&	2765		\\
Xylella fastidiosa 9a5c	&	xyfa	&	3 &NC\_002488	&	160492	&	2766 \\
\bf
Staphylococcus aureus NCTC 8325	&	stau\_	&	6 &NC\_007795	&	93061	&	2767 \\
Listeria monocytogenes EGD-e	&	limo	& 6 &	NC\_003210.1	&	169963	&	2867 \\
Synechocystis sp. PCC 6803	&	sysp	& 13 &	NC\_000911.1	&	1148	&	3179 \\
\bf
Burkholderia thailandensis E264	&	buth	&	2 & NC\_007651	&	271848	&	3276 \\
Sinorhizobium meliloti 1021	&	sime	& 1 &	NC\_003047.1	&	266834	&	3359 \\
\bf
Burkholderia pseudomallei K96243	&	bups	&	3 & NC\_006350	&	272560	&	3398 \\
Ralstonia solanacearum GMI1000	&	raso	&	2 & NC\_003295.1	&	267608	&	3436 \\
Clostridium acetobutylicum ATCC 824	&	clac	& 8 &	NC\_003030.1	&	272562	&	3602 \\
\bf
Caulobacter crescentus	&	cacr	&	1 & NC\_011916	&	565050	&	3885 \\
\bf
Mycobacterium tuberculosis H37Rv	&	mytu	&	5 & NC\_000962.3	&	83332	&	3936 \\
\bf
Escherichia Coli K-12 MG1655	&	esco	&	3 & NC\_000913.3	&	511145	&	4004 \\
\bf
Shewanella oneidensis MR-1	&	shon	&	3 & NC\_004347	&	211586	&	4065 \\
\bf	
Bacillus subtilis 168	&	basu	&	6 & NC\_000964	&	224308	&	4175 \\
\bf
Salmonella enterica serovar Typhi	&	saen	&	3 & NC\_004631	&	209261	&	4352 \\
\bf	
Bacteroides thetaiotaomicron VPI-5482	&	bath	&	7 & NC\_004663	&	226186	&	4778 \\
\bf
Sphingomonas wittichii RW1	&	spwi	& 1& 	NC\_009511	&	392499	&	4850 \\
\bf
Pseudomonas aeruginosa UCBPP-PA14	&	psae	&	3 & NC\_008463	&	208963	&	5892 \\
Mesorhizobium loti MAFF303099	&	melo	&	1 & NC\_002678.2	&	266835	&	6743 \\
Rickettsia prowazekii str. Madrid E	&	ripr	&	1 & NC\_000963.1	&	272947	&	8433 \\
\hline
\end{tabular}
\caption{Summary of the selected bacterial dataset. Organism name, abbreviation, class, RefSeq, STRING code, size of genome (number of genes $n$). 
Genomes annotated in the Database of Essential Genes (DEG) are highlighted with bold fonts. Classes are:Alphaproteobacteria(1), Betaproteobacteria(2), Gammaproteobacteria(3), Epsilonproteobacteria(4), Actinobacteria(5), Bacilli(6), Bacteroidetes(7), Clostridia(8), Deinococci(9), Mollicutes(10), Spirochaetales(11), Aquificae(12), Cyanobacteria(13), Chlamydiae(14), Fusobacteria(15), Thermotoga(16). }
\label{tab.dataset}
\end{table*}

%TABLE 2(tab.func)
\begin{table*}[h!]
\centering
\begin{tabular}{cl} 
\hline
{\bf COG ID}& {\bf Functional classification} \\
\hline
& {\em INFORMATION STORAGE AND PROCESSING}\\
%\hline
J & Translation, ribosomal structure and biogenesis \\
K &	Transcription \\ 
L & Replication, recombination and repair\\ 
\hline
&  {\em CELLULAR PROCESSES AND SIGNALING}\\
%\hline
D &	Cell cycle control, cell division, chromosome partitioning \\ 
T & Signal transduction mechanisms	\\
M &	Cell wall/membrane/envelope biogenesis \\
N & Cell motility\\
O & Post-translational modification, protein turnover, chaperones \\ 
\hline
& {\em METABOLISM}\\
%\hline
C & Energy production and conversion \\ 
G & Carbohydrate transport and metabolism \\
E &	Amino acid transport and metabolism \\ 
F & Nucleotide transport and metabolism \\
H & Coenzyme transport and metabolism	\\
I & Lipid transport and metabolism \\
P &	Inorganic ion transport and metabolism \\
\hline\end{tabular}
\caption{Functional classification of COG clusters.}
\label{tab.func}
\end{table*}

%TABLE 3 (tab.rete)
\begin{table*}[h!]
\centering
\begin{tabular}{c|ccc} 
\hline						
{\bf Organisms}	&	{\bf $r_{ee}$}	&	{\bf $r_{\bar{e}\bar{e}}$} & {\bf $r_{e\bar{e}}$} 	\\
\hline
basu	& 44.46& 	0.80& 	0.11 \\
bath	& 20.07 & 	0.76 & 	0.25 \\
bups & 	6.21	& 0.83 & 	0.27 \\
buth & 	18.69 & 	0.70	& 0.22 \\
cacr & 	18.40	& 0.70 & 	0.15 \\
caje & 	3.65 & 	0.82	&  0.32 \\
esco & 	2.91	& 0.88	&  0.31 \\
frno & 	9.84 & 	0.52 & 	0.18 \\
hain & 	1.65	&  1.15 & 	0.27 \\
hepy & 	2.91	&  0.78 & 	0.38 \\
myge & 	1.42	& 0.29 & 	0.08 \\
mypu	&  3.42	&  0.22& 	0.12 \\
mytu	&  8.09	&  0.78 & 	0.23 \\
pogi	&  11.03 & 	0.41	 & 0.21 \\
psae	 & 9.85 & 	0.92	& 0.16 \\
saen & 	28.80	 & 0.81 & 	0.12 \\
shon & 	6.50 & 	0.64	&  0.16 \\
spwi & 	15.47	&  0.74 & 	0.22 \\
stau & 	23.05 & 	0.58	 & 0.23 \\
stau\_ & 	21.89	&  0.64 & 	0.16 \\
stpy	 & 9.30	& 0.73 & 	0.23 \\
stsa	& 30.65	&  0.61 & 	0.22 \\
vich	& 8.37 & 	0.81	& 0.19 \\													
\hline
\end{tabular}
\caption{Relative density values $r$ for PPI subnetworks between essential genes ($r_{ee}$), between nonessential genes ($r_{\bar{e}\bar{e}}$) 
and between essential and nonessential genes ($r_{e\bar{e}}$), for each DEG-annotated bacterial genome.}
\label{tab.rete}
\end{table*}

%TABLE 4 (tab.hubs)
\begin{table*}[h!]
\centering
\begin{tabular}{c|c|c|c} 
\hline						
{\bf $k$}&{\bf COG}&{\bf Gene} & {\bf Protein}\\
\hline
57	&	1250I	&	paaH&	3-hydroxyadipyl-CoA dehydrogenase, NADdependent\\
	&	0365I	&	acs	&	acetyl-CoA synthetase	\\
58	&	0222J	&	rplL	&	50S ribosomal subunit protein L7$/$L12	\\
	&	0335J	&	rplS	&	50S ribosomal subunit protein L19	\\
	&	0267J	&	rpmG	&	50S ribosomal subunit protein L33	\\
	&	0365I	&	acs	&	acetyl-CoA synthetase	\\
59	&	0183I	&	paaJ	&	3-oxoadipyl-CoA3-oxo-5,6-dehydrosuberyl-CoA thiolase	\\
	&	1960I	&	ydiO	&	putative acyl-CoA dehydrogenase	\\
	&	0183I	&	atoB	&	acetyl-CoA acetyltransferase	\\
60	&	0197J	&	rplP	&	50S ribosomal subunit protein L16	\\
	&	0088J	&	rplD	&	50S ribosomal subunit protein L4	\\
	&	0197J	&	rplP	&	50S ribosomal subunit protein L16	\\
	&	0087J	&	rplC	&	50S ribosomal subunit protein L3	\\
	&	1960I	&	aidB	&	putative acyl-CoA dehydrogenase 	\\
	&		&		&		\\
61	&	0085K	&	rpoB	&	RNA polymerase, beta subunit	\\
	&	0202K	&	rpoA	&	RNA polymerase, alpha subunit	 \\
&		&		&		\\
62 & 0087J & rplC  & 50S ribosomal subunit protein L3 \\
& 0052J &  rpsB & 30S ribosomal subunit protein S2  \\
& 2965L & PriB  & ribosomal replication protein  \\
\hline
\end{tabular}
\caption{Specifics of the hub proteins that populate the few bins of connectivity around $k=60$ in Figure \ref{fig6}.}
\label{tab.hubs}
\end{table*}

%TABLE 5 MOVED TO SI (AG)

%TABLE 6 (tab.COGJ)
\begin{table*}[h!]
\centering
\begin{tabular}{lcccc}
\hline						
{\bf COG}&{\bf Genes name} & {\bf $<$k $>$}\\
\hline
COG0097J	&	50S ribosomal protein L6	&	60.24	\\
COG0087J	&	50S ribosomal protein L3	&	60.19	\\
COG0197J	&	50S ribosomal protein L16	&	60.19	\\
COG0090J	&	50S ribosomal protein L2	&	60.14	\\
COG0080J	&	50S ribosomal protein L11	&	60.12	\\
COG0088J	&	50S ribosomal protein L4	&	60.12	\\
COG0081J	&	50S ribosomal protein L1	&	58.19	\\
COG0089J	&	50S ribosomal protein L23	&	57.88	\\
COG0102J	&	50S ribosomal protein L13	&	57.45	\\
COG0094J	&	50S ribosomal protein L5	&	57.21	\\
COG0092J	&	30S ribosomal protein S3	&	57.12	\\
COG0098J	&	30s ribosomal protein S5	&	57.10	\\
COG0093J	&	50S ribosomal protein L14	&	57.00	\\
COG0091J	&	50S ribosomal protein L22	&	56.24	\\
COG0049J	&	30S ribosomal protein S7	&	55.31	\\
COG0051J	&	30S ribosomal protein S10	&	55.24	\\
COG0200J	&	50S ribosomal protein L15	&	55.12	\\
COG0256J	&	50S ribosomal protein L18	&	54.86	\\
COG0203J	&	50S ribosomal protein L17	&	54.43	\\
COG0244J	&	50S ribosomal Protein L10	&	54.19	\\
COG0100J	&	30S ribosomal protein S11	&	53.76	\\
COG0522J	&	30S ribosomal protein S4	&	53.43	\\
COG0096J	&	30S ribosomal protein S8	&	53.10	\\
COG0099J	&	30S ribosomal protein S13	&	52.88	\\
COG0048J	&	30S ribosomal protein S12	&	52.14	\\
COG0198J	&	50S ribosomal protein L24	&	50.83	\\
COG0185J	&	30S ribosomal protein S19	&	50.52	\\
COG0199J	&	30S ribosomal protein S14	&	50.45	\\
COG0103J	&	30S ribosomal protein S9	&	49.45	\\
COG0480J	&	tetracycline resistance protein. tetM &	47.90	\\
COG0052J	&	30S ribosomal protein S2	&	47.69	\\
COG0184J	&	30S ribosomal protein S15	&	45.95	\\
COG0186J	&	30S ribosomal protein S17	&	44.60	\\
COG0255J	&	50S ribosomal protein L29	&	43.95	\\
COG0222J	&	50S ribosomal protein L7/L12	&	42.43	\\
COG1841J	&	50S ribosomal protein L30	&	40.71	\\
\hline
\end{tabular}
\caption{Genes belonging to COG J with average degree bigger than 40 (see Figure \ref{fig7}). All these genes are conserved, common to all species (ERI=1), and drive the transition shown in Figure \ref{fig6}}.
\label{tab6}
\end{table*}

\begin{table*}[h!]
\begin{tabular}{|c|c|c|c|c|c|c|c|c|c|c} 
\hline						
{\bf k}	&	{\bf a}	&	{\bf b}	&	{\bf c}	&	{\bf d}	&	 {\bf $R^2$}	\\
\hline
$[0-10]$ & 40,561	&	-10,338	&	5,555	&	1,052	&	0,617	\\
$[10-20]$ & 23,774	&	3,890	&	8,583	&	0,626	&	0,590	\\
$[20-30]$ & 20,280	&	8,287	&	8,276	&	0,507	&	0,540	\\
$[30-40]$ & 18,296	&	10,685	&	8,334	&	0,190	&	0,790	\\
$[40-50]$ & 18,548	&	10,372	&	8,326	&	0,495	&	0,589	\\
$[50-60]$ & 25,868	&	2,508	&	8,038	&	0,650	&	0,758	\\
$[60-70]$ & 29,344	&	-0,756	&	10,224	&	0,977	&	0,870	\\
$[70-80]$ & 30,507	&	3,438	&	6,990	&	0,811	&	0,874	\\
\hline
\end{tabular}
\caption{ Best fit values of the parameters in equation (2) and correlation coefficients for different connectivity data, shown in figure \ref{fig4}.}
\label{tab5}
\end{table*}

\end{document}